\documentclass[twocolumn,showpacs,preprintnumbers,amsmath,amssymb,prl]{revtex4}

\usepackage[dvips]{graphicx}
\usepackage{dcolumn}
\usepackage[normalem]{ulem}

\usepackage{amsmath}
\usepackage{amssymb}
\usepackage{upgreek}
\usepackage{psfrag}
\usepackage{color}

\newcommand{\bm}[1]{\boldsymbol{\mathbf{#1}}}
\newcommand{\ud}{\mathrm{d}}

\newcommand{\bra}{\left\langle}
\newcommand{\ket}{\right\rangle}

\begin{document}

\preprint{}

\title{Subwavelength focusing inside an open disordered medium \\ by time reversal at a single point antenna}

\author{Romain Pierrat}
\email{romain.pierrat@espci.fr}
\author{C\'edric Vandenbem}
\author{Mathias Fink}
\email{mathias.fink@espci.fr}
\author{R\'emi Carminati}
\email{remi.carminati@espci.fr}
\affiliation{Institut Langevin, ESPCI ParisTech, CNRS, 1 rue Jussieu, 75238 Paris Cedex 05, France}

\date{\today}

\begin{abstract}
   We study theoretically light focusing at subwavelength scale inside a disordered strongly scattering open medium. We show that
   broadband time reversal at a single point antenna, in conjunction with near-field interactions and multiple scattering, produces spatial focusing 
   with a quality comparable to that obtained in an ideal closed cavity. This gives new perspectives for super-resolved optical imaging and 
   coherent control of single nanosources or absorbers in complex media.
\end{abstract}


\maketitle


The ability to focus light in a small region of space underlies many optical imaging techniques, including the most recent 
improvements in optical microscopy~\cite{HELL-2007}. Focusing beyond the diffraction limit is a major objective,
for the improvement of spatial resolution. One possible approach is offered by scanning near-field optical microscopy, 
in which a sharp tip produces a subwavelength spot
that can be scanned in the near field of the medium under investigation~\cite{NOVOTNY-2006}. Recently, the combination
of wavefront shaping and scattering by a structured environment has been recognized as an alternative way to produce
subwavelength focal spots in optics~\cite{MOSK-2010,MOSK-2011,SENTENAC-2008,STOCKMAN-2008}. 
An efficient control of wave propagation in complex media, including focusing, can be achieved by time reversal.
The technique is well established in acoustics~\cite{FINK-1997}, and has been extended to seismology~\cite{LARMAT-2006},
microwaves~\cite{LEROSEY-2004}, and to optics~\cite{VELLEKOOP-2007,CARMINATI-2010-4}.
The field of subwavelength light focusing encompasses a number of fields beyond imaging,
such as coherent control of single emitters or absorbers in complex media~\cite{CAO-2010-1}
or addressing in integrated optics~\cite{STOCKMAN-2008,LAGENDIJK-2011-1,STOCKMAN-2007,CAO-2011}.

A typical time-reversal focusing experiment consists in two steps. In the forward problem, the field radiated by a (point) source
in an arbitrary medium is recorded on an array of detectors (often called a Time-Reversal Mirror or TRM). For the time-reversed 
process, the detectors become sources sending back the time-reversed sequence of the recorded field. The result is a time-reversed field 
that focuses towards the source location. Time reversal below the diffraction limit has been demonstrated for the first time in acoustics, using 
an acoustic sink (i.e. an active time-reversed source) placed at the focal point~\cite{ROSNY-2002}. In a structured environment,
subwavelength focusing can be achieved even without creating a sink due to the presence of scatterers near the focal
spot~\cite{LEROSEY-2007}. Since recent technical progress has made possible the use of time reversal concepts for spatial (and temporal) 
focusing of light in complex media~\cite{FINK-2010-2,FINK-2012,LEROSEY-2012}, it seems that some fundamental questions need to
be examined: In particular, the feasibility of subwavelength focusing by time-reversal in an open disordered medium
with a single antenna has never been demonstrated, neither on the theoretical nor experimental
side. The goal of this Letter is to clarify this issue by analyzing
precisely the key role of the subwavelength disorder (multiple scattering, near-field interactions).


In this Letter, we address these questions theoretically based on numerical experiments. We show that subwavelength focusing in a strongly scattering
disordered open system is feasible by time reversal at a single point antenna. We put forward the crucial role of the spectral bandwidth,
in conjunction with near-field interactions and multiple scattering, to reach focusing performances comparable to that produced in an ideal closed cavity,
and that could not be achieved by monochromatic phase conjugation~\cite{RAHMANI-2010-2}.

\begin{figure}[!htbf]
   \centering
   \psfrag{a}[c]{(a)}
   \psfrag{b}[c]{(b)}
   \psfrag{R}[c]{$2R$}
   \psfrag{d}[c]{$\bm{r}_d$}
   \includegraphics[width=\linewidth]{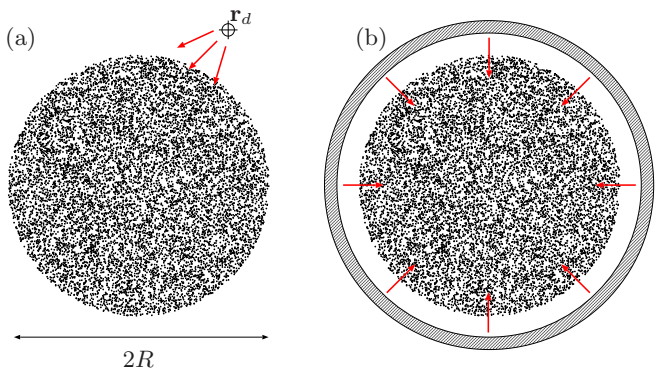}
   \caption{System composed by a 2D cluster of cylindrical scatterers randomly placed inside a cylinder or radius~$R$
   centered at position $\bm{r}_0$ where the source is lying in the forward problem (surrounded by an exclusion surface of
   radius $R_0$).
   Detection wth a single point sensor lying at position~$\bm{r}_d$ in the far field [subfigure~(a)] or a closed cavity
   (TRM) [subfigure~(b)]. Real scatterers positions are displayed on that figure.}
   \label{system}
\end{figure}

We consider a two-dimensional cluster of cylindrical scatterers randomly distributed inside
a cylindrical region of radius $R=1.91\,\upmu\textrm{m}$. One specific configuration of the cluster is used throughout the paper [Fig.~\ref{system}~(a)].
A minimum distance $d_{\textrm{min}}=10\,\textrm{nm}$ is forced between scatterers to avoid overlapping.
The scatterers are described by their electric polarizability
$\alpha\left(\omega\right)=-4\Gamma c_0^2/\left[\omega_0\left(\omega^2-\omega_0^2+i\Gamma\omega^2/\omega_0\right)\right]$
here $\omega_0=3.14\times 10^{15}\,\textrm{s}^{-1}$ is the resonance frequency, $\Gamma=10^{14}\,\textrm{s}^{-1}$ is the linewidth and
$c_0$ is the speed of light in vacuum. This corresponds to the Transverse Electric (TE) polarizability of a resonant non-absorbing
2D-point scatterer, with a quality factor $Q=\omega_0/\Gamma=31.4$. This very general form~\cite{LAGENDIJK-1998} of polarizability can also be applied
to a cylinder of given permittivity $\epsilon$ and radius $a$. In that case $\Gamma$ depends on $\epsilon$ and $a$.

The time reversal process is assumed broadband, in a spectral interval $\left[\Omega-\Delta,\Omega+\Delta\right]$, with
 $\Omega$ the central or reference frequency and $2\Delta$ the bandwidth. 
For the numerical simulations, we have chosen $\Omega=\omega_0-4\Gamma$ and $\Delta=2\Gamma$.
This corresponds to a reference wavelength $\lambda\left(\Omega\right)=688\,\textrm{nm}$.
The number of scatterers is $N=11683$ such that the average distance between scatterers is $\bra d\ket=31\,\textrm{nm}$.
At the reference frequency $\Omega$, the Boltzmann mean-free path is
$\ell_B\left(\Omega\right)=\left[\rho\sigma_s\left(\Omega\right)\right]^{-1}=219\,\textrm{nm}$ with
$\sigma_s\left(\omega\right)=(\Omega/c_0)^3/4\left|\alpha\left(\Omega\right)\right|^2$ the scattering cross-section and $\rho$
the density of scatterers. The optical thickness is $b=2R/\ell_B\left(\Omega\right)=17.4$, large enough
for multiple scattering to take place and $\bra d\ket/\lambda=0.045$, small enough for near-field interactions to occur. The
scattering strength is $(\Omega/c_0)\ell_B\left(\Omega\right)=2$.

In the forward problem, the 2D system is illuminated using a point source [dipole moment $p_0\left(\omega\right)=\operatorname{cte}$] 
polarized along the scatterers (TE modes), such that the electromagnetic problem is scalar. The source lies at the center of the cluster
denoted by $\bm{r}_0$ and is surrounded by an exclusion domain
with radius $R_0=10\,\textrm{nm}$, small enough to preserve near-field interactions between the source and the surrounding scatterers.
A single point antenna lies in the far field, outside of the system, at position $\bm{r}_d=\left(4,4\right)\,\upmu\textrm{m}$.
To solve Maxwell's equations and to compute the electric field at the position of the antenna, we use the Green
function $G$ that connects the electric dipole at position $\bm{r}_0$ to the radiated electric field at position $\bm{r}_d$
through the relation $E(\bm{r}_d,\omega) = \mu_0\omega^2G(\bm{r}_d,\bm{r}_0,\omega)p_0$. 
To proceed, we perform a coupled-dipole numerical computation.
The field exciting scatterer number $j$ is given by the contribution of the dipole source and of all other scatterers,
leading to a set of $N$ self-consistent equations~\cite{LAX-1952,FROUFE-2007}:
\begin{equation}\nonumber
   E_j=\mu_0\omega^2G_0\left(\bm{r}_j,\bm{r}_0,\omega\right)p_0
   +\alpha\left(\omega\right)k^2\sum_{\substack{k=1\\k\ne j}}^NG_0\left(\bm{r}_j,\bm{r}_k,\omega\right)E_k
\end{equation}
where $\bm{r}_j$ is the position of scatterer number $j$. 
$G_0\left(\bm{r},\bm{r}',\omega\right)=i/4\,\operatorname{H}_0^{(1)}\left(k \left|\bm{r}-\bm{r}'\right|\right)$ is the free space
Green function with $\operatorname{H}_0^{(1)}$ the Hankel function of first kind and order zero.
This linear system is solved numerically for the configuration of Fig.~\ref{system}~(a). Once the exciting electric field
on each scatterer is known, it is possible to compute the scattered field at any position and in particular
at the position of the antenna. In this numerical approach,
near-field and far field dipole-dipole interactions and multiple scattering are taken into account rigorously.

In the time-reversed process, the point source at position $\bm{r}_0$ is removed and the antenna becomes an active dipole source
with an amplitude $p_d\left(\omega\right)$ proportional to the complex conjugate of the recorded field. 
The same procedure as for the forward problem is used to compute the time-reversed field
$E_{\textrm{TR}}\left(\bm{r},\omega\right)$ in the vicinity of $\bm{r}_0$.


\begin{figure}[!htbf]
   \centering
   \includegraphics[width=\linewidth]{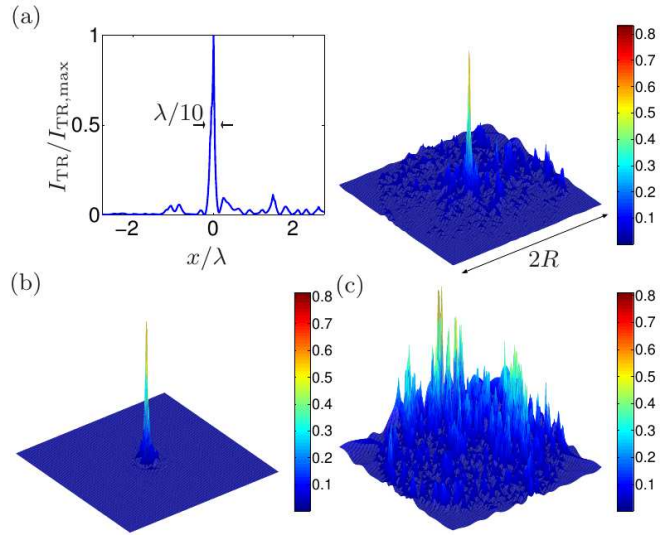}
   \caption{(Color online) Normalized broadband time-reversed intensity ($\Delta/\Gamma=2$) as a function of space for a
   single point antenna [subfigure~(a)] (the left figure is a section view of the right 3D plot) and for a closed cavity
   [subfigure~(b)]. Normalized monochromatic time-reversed
   intensity ($\Delta/\Gamma=0$) as a function of space for a single point antenna [subfigure~(c)].}
   \label{focal_spot}
\end{figure}

Figure~\ref{focal_spot}~(a) displays the time-reversed intensity integrated over the
frequency range $\left[\Omega-\Delta,\Omega+\Delta\right]$, defined as
\begin{equation}\label{intens}
   I_{\textrm{TR}}\left(\bm{r}\right)=\left|\int_{\Omega-\Delta}^{\Omega+\Delta}E_{\textrm{TR}}\left(\bm{r},\omega\right)\ud\omega\right|^2.
\end{equation}
We clearly distinguish a focal spot centered at $\bm{r}_0$, the position of the source in the forward problem. The width of the focal peak can be estimated
at $\lambda/10$ (full width at half maximum). This example demonstrates the feasibility of subwavelength focusing in an open strongly scattering medium
using a single point antenna. In order to determine the key parameters that explain the focusing process, we have performed
numerical experiments.

In a first step, we have replaced the single point antenna 
by an array of detectors placed in the far field forming a closed cavity around the system as shown in Fig.~\ref{system}~(b), in analogy with the
concept of TRM developed for acoustics~\cite{FINK-1997}. In such an ideal situation,
the time-reversed field is given by the imaginary part of the Green function of the scattering medium~\cite{CARMINATI-2007}:
\begin{equation}
   E_{\textrm{TR}}\left(\bm{r},\omega\right)=-2i\mu_0\omega^2\Im\left[G\left(\bm{r},\bm{r}_0,\omega\right)\right]p_0^*.
   \label{eq:ImG}
\end{equation}
The corresponding time-reversed intensity, computed using Eq.~(\ref{intens}) is plotted in Fig.~\ref{focal_spot}~(b).
The pattern is similar to that in Fig.~\ref{focal_spot}~(a), with a width of the focal spot still on the order of $\lambda/10$. 
We therefore conclude that time reversal at a single point antenna is here as efficient as time reversal in a closed cavity in terms of 
focusing performances. This is a consequence of strong multiple scattering in the disordered medium, that even in the case of an open
system creates conditions for time reversal that are very close to that known for closed chaotic cavities~\cite{FINK-2000}.

Nevertheless, strong multiple scattering and disorder are not enough to ensure focusing with super-resolution. In a second step, we have performed
similar simulations with an exclusion domain of radius $R_0=\lambda$ around the focal point $\bm{r}_0$ (not shown in this Letter for the sake of
brevity). This exclusion domain prevents near-field interactions between the source and the scatterers to occur. A time-reversed focal peak
is still visible in these conditions, but with a diffraction limited width of $\lambda/2$. This shows that near-field interactions with the surrounding
scatterers is necessary in order to generate a subwavelength concentration of light in the focus region. This mechanism has already been 
put forward in the context of time reversal of microwaves~\cite{LEROSEY-2007}. It is interesting to note that the same near-field interactions
are at the origin of the so-called $C_0$ correlation in speckle patterns produce by a point source in a disordered 
medium~\cite{SHAPIRO-1999,SAPIENZA-2011}. Since this correlation is connected to the fluctuations of the local density of 
states~\cite{TIGGELEN-2006,CAZE-2010}, the latter being given by an expression similar to Eq.~(\ref{eq:ImG}) with $\bm{r}=\bm{r}_0$,
a close connection with time reversal focusing might be established (that is beyond the scope of this Letter).

A crucial feature of time reversal is the spectral bandwidth. In a third step, we have studied the quality of focusing versus the bandwidth
measured by the parameter $\Delta$. In Fig.~\ref{focal_spot}~(c), we show the time-reversed intensity
$I_{\textrm{TR}}\left(\bm{r},\Omega\right)=\left|E_{\textrm{TR}}\left(\bm{r},\Omega\right)\right|^2$
 for time reversal at a single point antenna and at the single frequency $\Omega$ (which is equivalent to monochromatic phase conjugation). 
 The signal-to-background ratio is very weak and no focal spot emerges in the intensity map. This demonstrates the crucial role of
wideband time reversal for focusing in one realization of a disordered medium.
A key question is the determination of the optimal bandwidth. Intuitively, it should be determined by the spectral correlation
(or coherence) bandwidth $\delta_{\textrm{coh}}$ of the speckle pattern produced in the disordered medium. 
Let us define the spectral field-field correlation at point $\bm{r}_0$ as
(the brackets $\bra\cdots\ket$ denote the average over an ensemble of configurations of the disordered medium)
\begin{equation}\label{correl}
      C\left(\delta\right)=\frac{\bra E\left(\bm{r}_0,\Omega-\delta\right) E^*\left(\bm{r}_0,\Omega+\delta\right)\ket}
         {\bra I\left(\bm{r}_0,\Omega-\delta\right)\ket}
\end{equation}
and the coherence bandwidth $\delta_{\textrm{coh}}$ as the width of this correlation function.
A self-averaging process can be expected when $\Delta \gg \delta_{\textrm{coh}}$, giving a criterion for the bandwidth of the
time reversal process ensuring focusing in one single realization of the medium.
To compute $C\left(\delta\right)$, we have generated numerically $n=480$
random configurations as that in Fig.~\ref{system}~(a), and calculated the field at $\bm{r}_0$ for plane-wave
illumination. We have obtained a coherence bandwidth $\delta_{\textrm{coh}}\simeq 0.04\Gamma$, fifty times smaller than the 
time-reversal bandwidth $\Delta=2\Gamma$. This amounts to manipulating $\Delta/\delta_{\textrm{coh}}\simeq 50$
spectral degrees of freedom (compared to 1 in the monochromatic case), explaining the improvement in the
averaging process and the signal-to-background ratio.

\begin{figure*}[!htbf]
   \centering
   \psfrag{a}[c]{\textcolor{black}{(a)}}
   \psfrag{b}[c]{\textcolor{black}{(b)}}
   \psfrag{c}[c]{\textcolor{black}{(c)}}
   \psfrag{d}[c]{\textcolor{black}{(d)}}
   \psfrag{e}[c]{\textcolor{black}{(e)}}
   \psfrag{f}[c]{\textcolor{black}{(f)}}
   \psfrag{x}[c]{$x/\lambda$}
   \psfrag{y}[c]{$y/\lambda$}
   \includegraphics[width=0.95\linewidth]{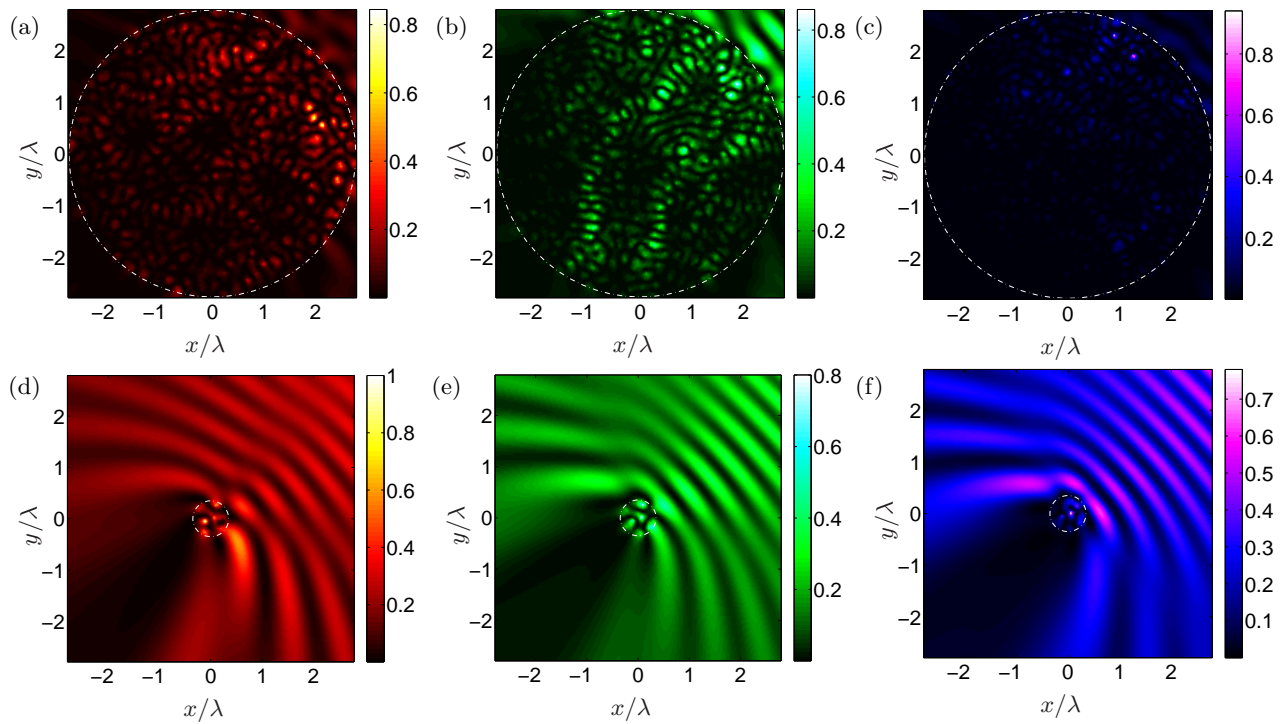}
   \caption{(Color online) Normalized monochromatic time-reversed intensity $I_{\textrm{TR}}\left(\bm{r},\Omega+\delta\right)$.
   The first three subfigures correspond to the case of a system with a large optical thickness $b=2R/\ell_B\left(\Omega\right)=17.4$
   [i.e. the system shown in figure~\ref{system}~(a)] for three different frequencies given by $\delta/\Gamma=0$
   [subfigure~(a)], $\delta/\Gamma=0.5$ [subfigure~(b)] and $\delta/\Gamma=1$ [subfigure~(c)]. The last three
   subfigures correspond to the case of a system with an optical thickness close to unity $b=2R/\ell_B\left(\Omega\right)=2.1$ [i.e. a small section
   of the system shown in figure~\ref{system}~(a)] for the same three different frequencies [respectively subfigures~(d),
   (e) and (f)]. In all subfigures, the white dashed-dotted circle represents the boundary of the scattering cluster. The position
   of the single point antenna can be guessed in the top-right corner of the maps because of the presence of interferences fringes.}
   \label{speckles}
\end{figure*}

In order to illustrate the relevance of spectral correlations, we show in Fig.~\ref{speckles} maps of the time-reversed
intensity for different frequencies around the reference frequency $\Omega$, and for a system in the multiple scattering regime 
(top row, optical thickness $b=17.4$) or in the single scattering regime (bottom row, optical thickness $b=2.1$). 
The three maps in the multiple scattering regime [subfigures~(a), (b) and (c)] are clearly different, thus corresponding to three uncorrelated 
frequencies. Indeed, the coherence width in this case is $\delta_{\textrm{coh}} \simeq 0.04\Gamma$, small compared to the frequency steps between 
each map. In the single scattering regime [subfigures~(d), (e) and (f)], the three maps present some similarities. The coherence width is 
$\delta_{\textrm{coh}}\simeq 3\Gamma$, larger than the difference in frequencies in the maps that remain correlated.
This illustrates the positive role of multiple scattering in the efficiency of a broadband time-reversal process: a reduction
of the coherence width means a reduction of the bandwidth necessary to get self averaging in a single realization. An important conclusion
of this work is that this behavior is preserved for focusing at subwavelength scale in the presence of near-field interactions and strong disorder.


In summary, broadband time reversal at a single point antenna, in conjunction with near-field interactions and multiple scattering, is an
efficient technique for spatial focusing of light at subwavelength scale in a disordered medium. The time reversal 
bandwidth necessary for self averaging has been connected to the spectral correlation width of the speckle pattern in a regime that was
unexplored so far. These results should give new perspectives for super-resolved optical imaging in complex media, and for the 
coherent control of single nanosources, including quantum emitters, or nanoscale absorbers.
This work is supported by LABEX WIFI (Laboratory of Excellence within the French Program ``Investments for the Future'') under
reference ANR-10-IDEX-0001-02 PSL*.


\end{document}